# New thoughts on an old riddle: what determines genetic diversity within and between species?

Shi Huang

State Key Laboratory of Medical Genetics
School of Life Sciences
Xiangya Medical School
Central South University
110 Xiangya Road, Changsha, Hunan 410078, China

*Corresponding author: huangshi@sklmg.edu.cn

**Abstract**

The question of what determines genetic diversity both between and within species has long remained unsolved by the modern evolutionary theory (MET). However, it has not deterred researchers from producing interpretations of genetic diversity by using MET. We here examine the two key experimental observations of genetic diversity made in the 1960s, one between species and the other within a population of a species, that directly contributed to the development of MET. The interpretations of these observations as well as the assumptions by MET are widely known to be inadequate. We review the recent progress of an alternative framework, the maximum genetic diversity (MGD) hypothesis, that uses axioms and natural selection to explain the vast majority of genetic diversity as being at optimum equilibrium that is largely determined by organismal complexity. The MGD hypothesis fully absorbs the proven virtues of MET and considers its assumptions relevant only to a much more limited scope. This new synthesis has accounted for the much overlooked phenomenon of progression towards higher complexity, and more importantly, been instrumental in directing productive research into both evolutionary and biomedical problems.

## Introduction

The modern evolutionary theory (MET) consists of Darwin's theory of natural selection and Kimura's Neutral theory (also Ohta's Nearly Neutral theory). The theory treats evolution the same as population genetics. The Darwinian theory is much better known than the Neutral theory. However, for molecular evolution and population genetics, the Neutral theory (and the Nearly Neutral theory) has been more useful. Regardless, however, the MET is still incomplete, as acknowledged by Ohta and Gillespie: "..we have yet to find a mechanistic theory of molecular evolution that can readily account for all of the phenomenology. ..we would like to call attention to a looming crisis as theoretical investigations lag behind the phenomenology." [1].

Key puzzles of evolution remain unsolved by the MET. The central problem of the field is and has always been the old riddle of what determines genetic diversity [2-5]. Is it mostly determined by natural selection or neutral drift? Here we critically examine the historical origins and assumptions of the MET. We show that both the neutral and the selection frameworks were largely mistaken right from the beginning. Key observations that directly inspired the neutral theory were nearly half of a century ahead of their time. Selection schemes on the other hand was largely influenced by the one gene one trait genetics of the early 1900s and always treated single locus as the target of selection, which is in fact rarely the case for most of the commonly observed variations as recent studies have shown [6-11]. Finally, we review a candidate for superseding the MET, the maximum genetic diversity (MGD) hypothesis first published in 2008 [12,13], that fully absorbs the proven virtues of the MET and has more explanatory power as well as greater value in directing productive research in a much wider field of biomedical science [6-11,14]. The old riddle of genetic diversity within and between species is solved as mere deductions of the assumptions of the MGD. Only in this case, the assumptions are, for the first time in biology, self-evident intuitions that are no less true or false than any known axioms of hard sciences or mathematics.

## Genetic diversity, a nearly century old riddle

Darwin's theory of common descent suggests a key role for variation and diversity in the diversification of new species from a common ancestor. The study of variation/diversity has a long history [15]. The advances in genetics in the early 1900s realized an important role of genetic variations in species diversification. Therefore, understanding the amount and nature of genetic variation has been fundamental to evolutionary research ever since the synthesis of natural selection with genetics in the 1930s and remains so today. For over two decades before 1960s, there were two schools of thoughts in hot debate, with one believing in very little genetic variation within a population of a species, and the other hypothesizing extensive genetic variation maintained by natural selection [4,5]. The debate was mostly fruitless because there was little experimental data on genetic variation. The question of what determines genetic difference between species is obviously directly related to within species variation since such variation is the seed for speciation, and as such was expectedly also unsolved. Diversity between different species in some cases could be considered as diversity within a clade or kind of species, e.g, diversity within the bacteria clade/kind.

In the mid-1960s, the application of protein electrophoresis method showed for *Drosophila*, humans and other organisms that there was extensive variation in many proteins [4,5]. The debate shifted to how to best explain such high diversity and the mysteriously narrow range of genetic diversity levels seen across taxa that vary markedly in their census population size [2]. There were again two main schools of thoughts that were naturally evolved from the earlier two schools, with the school originally believing in little genetic variation now favoring neutral drift, and the other sticking with natural selection. However, neither can fully account for the experimental observations, as Lewontin said of the neutral school: "…we are required to believe that higher organisms including man, mouse, Drosophila and the horseshoe crab all have population sizes within a factor of 4 of each other. …The patent absurdity of such a proposition is strong evidence against the neutralist explanation of observed heterozygosity." [5]. Regarding the selection school, the immense genetic load problem makes any simple natural selection schemes unrealistic, as pointed out by Lewontin and Hubby [16]. (However, as we will see later, this criticism of the selection school was based on the limited understanding of how selection may work and the mistaken view of the genetic load) The debate remains unsettled and has been neglected in the last three decades with the focus of the field shifted to generating more and more diversity data with newer techniques. Also going on heavily during this period is the free application of the neutral paradigm in interpreting observed genetic diversity data. Results from such analysis are routinely presented as if they are uncertainty free. Among the most notable and counter-intuitive results are the junk DNA notion and the Out of Africa model of human origins.

Research concerning neutral processes in genetics started in the 1920s and culminated in Kimura's neutral theory [17,18]. From the beginning researchers did not seek to deny the importance of natural selection but instead were interested in how neutral processes affected adaptive evolution [19]. Fisher and Wright asked questions and developed techniques of population genetics that are also relevant to the neutral theory [20,21]. Haldane's genetic load argument was instrumental in inspiring Kimura to develop his neutral theory [22]. In addition, some researchers have suggested that much of molecular evolution is neutral [23,24]. However, Kimura first combined population genetics theory with molecular evolution data to arrive at a theory of neutral evolution [17]. While this theory has since been applied to explain electrophoretic protein polymorphisms or diversity within species, the data that directly inspired Kimura was from sequence variation between species.

Just a few years earlier to the finding of electrophoretic protein polymorphism, genetic difference between species has been observed for hemoglobin, cytochrome C, and fibrinopeptides [25-27]. There were two kinds of results depending on whether one was comparing all concerned species to the apparently most complex among them such as human or the least complex such as yeast. If to human, one would observe 'the gradually increased amount of difference found when human hemoglobin is compared with hemoglobin from progressively more distant species' [25-27]. This directly inspired Zuckerkandl and Pauling to informally propose the universal molecular clock that hemoglobins from different species are changing at a steady and similar rate of $1.4 \times 10^{-7}$ amino acid substitutions per year [28].

If on the other hand, one compares human and frog to a lower complexity species like fish, one would observe a highly unusual result, the equidistance result that fish is approximately equidistant to human and frog in cytochrome C [26]. Similarly, yeast cytochrome C is equidistant to all multicellular organisms such as fish, frog, birds, horse, and human [26]. Perhaps because of the longtime influence by Darwin's theory that flatly ignores the issue of complexity, Margoliash, in first reporting the equidistance result of cytochrome C, only noted time to be the determining factor of genetic distance: "It appears that the number of residue differences between cytochrome c of any two species is mostly conditioned by the time elapsed since the lines of evolution leading to these two species originally diverged. ... If elapsed time is the main variable determining the number of accumulated substitutions, it should be possible to estimate roughly the period at which two lines of evolution leading to any two species diverged." [26]. With that, he formally proposed the universal molecular clock hypothesis that different species has roughly the same substitution rate, which implies that genetic distance between species is strictly determined by time alone.

In retrospect, Margoliash could as well have noticed the other striking feature and his sentence above could be rewritten as: "It appears that the number of residue differences between cytochrome c of any two species is mostly conditioned by *the species with lower organismal complexity.*" However, another sentence of Margoliash indicated clearly that he had no appreciation for function/complexity/physiology in constraining DNA variation: "It should be noted that the present results are compatible only with the commonly accepted scheme of evolution represented by series of branching lines, and are not consistent with a simultaneous formation of all species, which then proceed to accumulate mutations independently. In the latter case all the cytochromes c should be equally different from all others." [26]. It is obvious that the equidistance result was way ahead of its time. We now know that if human, fish and bacteria were created at the same time with identical cytochrome c, which then proceed to accumulate mutations, we would still find that human is closer to fish than to bacteria [12,13].

Although the two results from the above two kinds of alignment can both give rise to the idea of a universal molecular clock and are just two sides of the same coin, the equidistance result however was unusual for two reasons. First, unlike the distance to human result, deducing a molecular clock from the equidistance to a less complex species was more straightforward involving no calculations. Second, the distance to human result can be and has in fact been presented in textbooks as evidence for Darwin's theory. However, the equidistance result is completely unexpected: how can similar amounts of genetic change result in vastly different amounts of phenotypic change? This challenges the MET as the theory allows no role for epigenetics (epigenetic changes can drastically alter phenotypes without a change in DNA sequences). Despite being considered by some scientists from outside the field as 'one of the most astonishing findings of modern science', the genetic equidistance result has largely disappeared from our collective consciousness [29].

In addition to within-species diversity, key observations related to diversity among closely allied species and between distantly allied species, remain unexplained. The largest observed variation between different species of one clade is greater than that of another clade, which shows no correlation with time. Two rodent species separated no more than 13 million years ago show 2 fold greater genetic difference than between human and monkey (18% vs 8% in genomic DNA) that have separated for ~40 million years [30]. Two flowering plants (Arabidopsis and apple tree) that shared a common ancestor no more than 125 million years ago have more dissimilarity in DNA than humans and birds that shared a common ancestor 310 million years ago [12].

**Weaknesses of the modern evolution theory**

> The only real valuable thing is intuition.
> - Albert Einstein
> I don't believe in empirical science. I only believe in a priori truth.
> - Kurt Godel

Mathematicians/statisticians such as Fisher, Haldane, Wright, and Kimura have played an important role in developing the MET. Even pure mathematician and number theorist like Hardy has contributed a key equation to the field. It is the branch of biology with the most mathematics. But contrary to naive expectations, this field, even relative to other branches of biology such as biochemistry and molecular biology, is more like soft science than to hard core physics. In the words of an expert Jerry Coyne: "In science's pecking order, evolutionary biology lurks somewhere near the bottom, far closer to phrenology than to physics. For evolutionary biology is a historical science, laden with history's inevitable imponderables."[31].

Researchers of the field appear to prefer the self-serving notion that evolutionary studies are inherently more complex and involve impractical time scales. While that might be true, the real reason may be quite obvious and simple. Mathematics depends on assumptions or paradigms. The assumptions for hard core sciences are axioms or self-evident intuitions. Newton's axioms come to mind. They are all a priori true. In contrast, there is not a single assumption in the evolutionary field that is self-evidently true. Nearly all assumptions of the field are in fact self-evidently false, as Ohta and Gillespie said in 1996: "all current theoretical models suffer either from assumptions that are not quite realistic or from an inability to account readily for all phenomena." [1].

Below we critically examine the major assumptions of the field that were first set up long before people have much understanding of the genetic and phenotypic complexities. Most of these assumptions can be found in standard textbooks of molecular evolution [32-34].

*The infinite-site model.* Kimura and Crow formulated the infinite-allele model in 1964[35]. This model has been widely used for interpreting observed polymorphisms and for constructing phylogenetic trees. It is, however, obviously untrue as an organism's genome is finite in size and essentially nothing compared to infinity. Within such finite size genomes, the proportion that can be free to change without consequences is even more limited. A corollary of the infinite site assumption includes many unrealistic notions such as infinite genetic distance/diversity, no recurrent mutations, and no stage of evolution could be at an equilibrium. Under this model, genetic polymorphisms is simply "a transient phase of molecular evolution"[18]. However, real data suggests otherwise and genetic diversity observed today are in fact mostly at optimum equilibrium [6-11].

*Independent mutations.* Individual mutations are assumed to be independent events with independent effects on fitness. It is critical for deducing the genetic load argument. This assumption ignores the pervasive reality of epistatic effects of mutations.

*Fitness determined solely by genetic make-up and all loci contribute independently to fitness.* This assumption ignores the roles of epigenetics and the collective effects of all loci [6-11].

*Classification of mutations into deleterious, neutral and advantageous.* This assumption overlooks the fact that all mutations including advantageous ones have a deleterious aspect as random noises to an ordered system. It does not recognize the fact that most variations appear neutral as a result of balancing selections [6-11].

The above assumptions in effect assumed biological systems to be more like junk yards than a highly ordered network since only a junk yard can tolerate an infinite amount of building blocks and of errors in the building blocks where no epistasis exists for the errors. The junk DNA notion is only a deduction with zero experimental evidence and an illogical one for that matter because of the fallacy of begging the question [36]. The conclusion of junk DNA is already embedded in the assumptions that are used to deduce it.

*Random mating.* The random mating assumption is critical for many mathematical deductions in the field including the Hardy-Weinberg equation. However, most organisms simply do not engage in purely random mating. There are some randomness in selecting a mate but it is not completely random.

*Synonymous mutations are neutral.* This assumption is central to many commonly used tests for detecting natural selection such as the McDonald Krietman test [37]. However, evidence to the contrary has constantly appeared in recent years [38,39].

*Nonconservation means non-function.* This assumption treats many genomic elements as neutral such as repetitive and viral elements and is key to the conclusion of only 8% functional genome in humans [40]. However, to use conservation as an index of function is only measuring one of two kinds of sequences, the essential ones for internal system physiology that have little to do with adaption to the outside environments. To maintain the long term integrity of the system, such sequences cannot change. For living fossils to be possible, these sequences should be highly stable. On the other hand, sequences involved in adaption to environments must be fast changing because environmental changes are usually fast. Flu viruses escape neutralizing antibodies every few years, and the fast changing non-conserved sites in these viruses are absolutely critical for their survival but not essential for their physiology.

*Critical parameters of key equations not measured but tautologically derived.* Effective population size is deduced to be related to genetic diversity and often tautologically derived by using the very genetic variables that it is meant to predict or explain. This problem has been long ago pointed out by Lewontin [5].

The inadequacy of the MET for within species diversity has been well appreciated before and revisited recently [2]. Here, we discuss between species diversity and its misreading by the MET [41]. The constant mutation rate, i.e., molecular clock, interpretation of the first results of protein alignment is in fact a classic tautology, a mere ad hoc restatement of a distance phenomenon. It has not been verified by any independent observation and has on the contrary been contradicted by a large number of facts [12,33,42-51]. Nonetheless, people have treated the molecular clock as a genuine reality and have in turn proposed a number of theories to explain it [17,52-56]. The 'Neutral Theory' has become the favorite [17,55,56], even though it is widely acknowledged to be an incomplete explanation for the clock [45,57].

The abstract of the Kimura paper has only one sentence: "Calculating the rate of evolution in terms of nucleotide substitutions seems to give a value so high that many of the mutations involved must be neutral ones." But this calculation has two implicit assumptions that were taken for granted, without deliberation. One is that observed genetic distance always increases with time. The other is that every nucleotide in a genome is freely changeable (there are no nucleotide positions that would cause lethality when changed). As a matter of fact, the mere use of the equation r = d/2t for deriving the mutation rate is already guilty of begging the question: it has already assumed the conclusion, the same mutation rate, for the two species concerned when it assumes the distance to have been contributed by both equally.

The observed rate is measured in years but the Neutral theory predicts a constant rate per generation [41]. Also, the theory predicts that the clock will be a Poisson process, with equal mean and variance of mutation rate. Experimental data have shown that the variance is typically larger than the mean. Ohta's "nearly neutral theory" explained to some extent the generation time issue by observing that large populations have faster generation times and faster mutation rates, but remains unable to account for the great variance issue [58].

**The common sense approach to the old riddle**

As well said by Einstein: "Problems cannot be solved with the same mindset that created them." The difficulty of the MET in explaining the old riddle was created by a mindset long influenced by Darwin's theory that prefers to ignore the obvious reality of a direction of evolution from simple to complex. We as geneticists and cancer biologists, however, came to the field with a much more open mind not much burdened with any unrealistic assumptions.

In 2005, we independently re-discovered the genetic equidistance result and were naturally very shocked. We quickly realized the molecular clock interpretation as a mere tautology. In 2008, we published the maximum genetic diversity (MGD) hypothesis and reinterpreted all major evolutionary phenomena including the genetic equidistance result [12,13]. We also discovered a new characteristic of the equidistance result, the overlap feature, that has not even been appreciated let alone explained by any theory [59].

It is obvious that differences in biological complexity exist. People have since Aristotle long recognized the hierarchy in organismal complexity and order. Physics research into the cosmos also found that the universe evolves from simple to complex [60]. To study evolution must require an account of biological complexity.

The major building blocks for biological organisms are DNAs and the architectural plans of how to use the DNA parts are the epigenetic programs. A significant portion of epigenetic information may be located in the non-DNA materials of the egg. While certain mathematicians such as John von Neumann and Gregory Chaitin have considered the DNA code as software [61,62], modern findings of biology do not support such a notion. How can the same software produce dramatically different outputs such as different cell types within an organism? Not surprisingly, such naïve notion has not been useful in understanding genetic variation.

The more the cell types, the more the number of ways of using the same set of DNAs, and the more complex the organism [63-70]. Epigenetic programs are not only inherited during mitotic cell division but are also transmitted through the germline to the next generation, and sometimes over many generations [71-74]. We therefore define complex organisms as those that have complex epigenetic programs in terms of the number of cell types and epigenetic molecules. Furthermore, we assume that humans can in most cases intuitively and correctly judge the complexity differences of species. That human is more complex than monkey is an intuitive assumption that may not be possible to verify today but has no contradictions for now and is unlikely to be contradicted in the future.

With the above definition of complexity, we proposed the maximum genetic diversity (MGD) hypothesis based on a pair of intuitions or axioms [12,13,41]. Axiom 1 posits that the more complex the phenotype, the greater the restriction on the choice of molecular building blocks. In biology, this axiom means that there is an inverse relationship between genetic diversity and epigenetic complexity. Genetic diversity is defined here as genetic distance or dissimilarity in DNA or protein sequences between different individuals or species.

Axiom 2 says that any system can allow a limited level of random errors or noises in molecular building parts and such errors may be beneficial, deleterious, or neutral depending on circumstances. Limited errors at optimum level are more likely to be beneficial than deleterious because they are after all within tolerable levels and confer economy in construction and strongest possible adaptive capacity or robustness to environmental challenges. Obviously, one only needs to substitute "errors in building blocks" to "genetic diversity" to get the equivalent concept in biology. Axiom 2 in fact highlights the valid parts of MET as well as Nei's theory of niche-filling or mutation-driven evolution [75]. Speciation within a clade of similar complexity may proceed via either survival of the fittest or survival of the niche-filling variants. The idea of a limit on variation was first appreciated by the 'sphere of species variation' concept of Fleeming Jenkin [76,77].

We define macroevolution as an increase in organismal complexity, which can be mirrored by the increase in the precision of the building parts or a decrease in the allowed range of the standard deviations (stdev) for the parts (Fig. 1a). We define microevolution as an increase in genetic diversity within the allowed stdev ranges or as genetic turnovers at the equilibrium level of genetic diversity (Fig. 1b).

Genetic diversity cannot increase indefinitely with time and has a maximum limit being restricted by function or epigenetic complexity. The maximum genetic diversity of simple organisms is greater than that of complex organisms. Over long evolutionary time and for fast evolving DNAs, the genetic distance between sister species and a simpler outgroup taxon is mainly determined by the maximum genetic diversity of the simpler outgroup, although over short time scales it is mainly determined by time, drift, environmental selection, and the neutral mutation rates of the simpler outgroup as well as to a smaller extent by the rates of the sister taxa.

The MGD hypothesis explains the genetic equidistance phenomenon as a result of maximum genetic distance [12,13,41]. This phenomenon has in fact another characteristic, the overlap feature [59]. It shows a large number of overlapped mutant amino acid positions where any pair of any three species is different. Such positions indicate that at least two different species have each had independent mutations occurring at the same sites, which is unexpected if mutations occur randomly in a genome of infinite sites. These overlap positions therefore show a non-random nature in where mutations occur. They are however fully expected if mutations can only occur within a limited number of sites as defined by the range of tolerable errors and if time is long enough or mutation rates fast enough for all allowed sites to have mutated at least once in all species.

There are in fact two kinds of genetic equidistance result [12,13,44]. For long evolutionary timescale or for fast evolving sequences, one would observe "maximum genetic equidistance": different species are equidistant to a species of lower or equal complexity. The original result of Margoliash is maximum genetic equidistance. For short evolutionary timescale or for slow evolving sequences, one observes "linear genetic equidistance" where the molecular clock holds and distance is still linearly related to time: when sister species have similar mutation rates, they would be equidistant to a lower or equal complexity outgroup.

The two kinds of equidistance can be easily distinguished by the overlap feature. The observed number of overlap positions in cases of maximum genetic equidistance are consistent with the predictions of the MGD but far more than that by the neutral theory[59]. For linear genetic equidistance of microevolution, both theories gave similar predictions that match well with observations.

The so-called molecular clock is really about the constant rate of complexity increases. The first molecular evidence for constant albeit discontinuous advances in complexity is in fact the maximum genetic equidistance phenomenon. Defying the Darwinian gradualistic worldview, there have always been researchers since the time of Darwin who have appreciated the discontinuous nature in macro-evolutionary changes [15,29,76-79].

During evolution, genetic diversity, so long within a limit, should be the more the better, since it increases the adaptive capacity. Thus it would be positively selected and quickly reach MGD. Most parts of the genome have relatively fast mutation rates, and the observed genetic distances today are mostly maximum distances. For most species today, the variations in their genomes are mostly at an optimum equilibrium. The only exceptions are the slowest evolving parts of the genome [14].

The MGD predicts the existence of conserved sequences that are only related to complexity but not to enzyme function per se. The length of such sequences increases with complexity. This type of sequences has been found to exist [80], which has been termed Complexity Associated Protein Sectors or CAPS.

Functional constraint on sequence divergence has long been appreciated, including protein structure and function, genome-wide constraints such as fold pressure (nucleic acid stem-loop extrusion pressure) and GC-pressure (the pressure for a certain base composition), and purine loading pressure (purine rich mRNA synonymous strands), mtDNA and nuclear genome compatibility, Donnan equilibrium, aggregation pressure (crowded cytosol), and protein mass/volume/heat [18,77,81-84]. These constraints are thought to weaken the neutral theory, but their relationships with organismal complexity (and hence the equidistance phenomenon) remain unclear.

Having explained the diversity between species, we next discuss within-species diversity or the electrophoretic protein polymorphism and a few other puzzling observations of great interest.

*Less complex organisms show higher within species genetic diversity than complex ones.* A summary of this pattern can be seen in Leffler et al [2]. Obviously, such observation is a direct logical deduction of the MGD so long evolutionary time is long enough or the DNA being concerned has a fast enough mutation rate. Human has the lowest genetic diversity among all species is because of its high complexity rather than time.

*The mysteriously narrow range in within species genetic diversity.* This mystery has been highlighted recently [2]. Simple organisms such as bacteria are expected to have much greater MGD than primates. But the observed MGD of a specific type of bacteria species may not be that much greater than a specific type of monkey, which may seem inconsistent with the MGD hypothesis. However, if one looks at within clade between species diversity, one observe the expected results. The between species diversity within the bacteria clade is much greater than that within the primate clade. The bacteria kind has much greater MGD, which is one reason for the huge amount of diversification in bacteria species. However, during species diversification by the process of microevolution involving no change in MGD, certain sequences with adaptive roles to whatever environmental factors will become fixed as part of the speciation process, leading to a reduction in within species diversity. Others have recently found new evidence for natural selection in shaping genetic diversity [85]. However, without taking into account of the MGD concept, such attempts have only limited explanatory power.

*Genetic load.* One of most commonly used arguments for junk DNA is that if mutations are deleterious, the observed mutation rate would be too fast for human to tolerate such a high mutation load burden [36]. However, the genetic load concept is based on the untrue assumptions of independent mutations and of natural selection always working on single loci. If mutations are negatively selected in a collective way whenever a threshold of MGD is surpassed, the mutation load would never materialize. This is just like computers eventually crash due to long term use and the accumulation of all kinds of small random hits to its parts. Evidence for the deleterious nature of too much genetic noises has recently been found [6-11].

*If most common polymorphisms are under selection, how they appear to be neutral?* Most common SNPs have a distribution pattern consistent with the Hardy-Weinberg equation, which only a neutral SNP is expected to follow. That a SNP may appear to be neutral but is nevertheless under selection in a collective way is in fact fully expected. The collection of SNPs over a MGD threshold in an individual is largely accomplished through combination by mating, which has some random elements to it [6-11]. Thus selection on SNPs in a collective way is much more random than classical natural selection on a single locus [6-11]. Thus, SNPs under such random selective forces would be expected to behave similarly as truly neutral SNPs undergoing random drift.

*C-value paradox.* Genome sizes vary greatly and do not correlate with organismal complexity. Some plants like onion have much greater genome size than humans, which has been used to favor junk DNA idea [36]. However, houses can be small or large and so the number of parts involved can also vary from small to large. Genome as a whole is a building part, and this part can vary in size is just like any parts should have an allowable stdev. Along this line, it fully explains why complex mammals have much less variation in genome size relative to lower species such as fish or plants. The simple unicellular organism protozoa kind has a genome size variation range with one end ~20000 fold larger than the other end whereas mammals only less than 10 fold. The real significant question on genome size should be why the protozoa kind has such huge variation in genome size, rather than why the onion has such a huge genome.

## Is the MGD hypothesis useful?

A real understanding of the past should greatly help with understanding the present. Despite being only a few years old, the MGD hypothesis has been instrumental in directing productive research in both evolutionary problems and important biomedical problems. The hypothesis does not mean discarding the old assumptions but merely makes them more limited in their relevance. One must carefully select those DNAs that may follow those assumptions.

*Phylogenetics.* The MGD should help resolve difficult historical problems such as the phylogenetic tree of life. Past methods have no concept of maximum distance and used mostly non-informative distance data for inferring phylogeny. They often produce self-conflicting results and results inconsistent with the fossil records. We developed the slow clock method based on the MGD [14]. The method makes use of only slow evolving sequences that have zero overlap positions, and thus insures the linear relationship between distance and time. Its results therefore will be more objective and independent of the variations in sequence selections and investigators. Slow evolving sequences are more likely to meet the neutral criteria. They are unlikely to be under positive selection since their low speed of changes makes them too slow to meet adaptive needs. Relative to fast evolving sequences, they are also less likely to be under negative selection by way of collective effects over a MGD threshold since their low speed means that they are not a major contributor to the collective effects of variants. Also, such sequences are unlikely to be under pressure to reduce their tolerable number of changeable positions as a result of complexity increases, because their slow speed of mutations means that they are less likely to be disruptive to increased complexity. Thus, their MGD levels are more likely to be similar in different species. We have used the slow clock method to re-establish a correct primate phylogeny that re-establishes the intuitive common sense that human and pongids are two separate groups, which has long been the consensus view of the paleoanthropologists [14].

The notion that observed variations in slow evolving DNAs are neutral seems counter intuitive. However, it is in fact fully expected even from the old framework, as Lewontin said: "Thus (according to the neutral theory) the more "useless" a protein is physiologically, the more rapidly it should evolve. In contrast, a theory which holds that most of the gene substitution in evolution has been adaptive would predict that "useless" proteins would be among the slowest evolving."[5]. We would only correct him here that "useless" proteins may also be useful. A protein may be useless for adaptation but essential for internal system construction (examples may include house-keeping proteins such as actin). If variations are mostly for adaptive purposes in the sense of adapting to environments, then variations in those slow changing DNA sequences mostly responsible for system construction would be "useless" and hence neutral.

To such truly neutral sequences still at the linear phase of divergence, many of the assumptions of MET such as the infinite sites model and independent mutations would be valid. Thus phylogenetics research can largely proceed as before except that one has now a standard to separate the neutral from the noninformative DNAs. One must now distinguish two different kinds of high sequence similarity, one due to less time of separation and the other because of common construction resulting in using similar parts. We have shown that human and pongid are two separate clades diverged ~18 million years ago. The fact that human is closer to one than to the other of the species within the pongid clade (closer to chimpanzees than to orangutans) is due to one has more functional similarity (and hence the related similarity in building parts) with humans. Orangutan has less reasoning ability than chimpanzee and human [86]. In truly neutral slow evolving genes and when compared to orangutans, human is more distant while chimpanzee is closer, which hence justify a pongid clade and show a unity of molecules and phenotypes.

*Population genetics.* The present theoretical foundation for population genetics is the neutral and nearly neutral theory. The value of the theory is its description for the linear phase, which has been retained by the MGD. What the neutral theory lacks but the MGD provides is a consideration for the plateau phase of the evolutionary process when genetic distance no longer changes with time. We predict that genetic diversity of a typical population is mostly at an optimum level. Most of what is considered to be neutral variations by the neutral theory will turn out to be only seemingly neutral or neutral in the sense of the traditional Chinese 'The Way of the Neutral (middle/mean)', i.e., a net result of yin and yang selection. Recent studies inspired by the MGD framework have for the first time investigated the collective effects of variations or the effect of minor allele contents (MAC) in an individual [6-11]. They have shown that most minor alleles of SNPs are functional or under both negative and positive selection with negative selection slightly more dominant.

*Major biomedical problems.* Most complex traits and diseases are partly inheritable and presumably caused by polymorphic genetic variations such as SNPs. The neutral theory views most such variations to be nonfunctional and neutral and hence the study of complex traits and diseases has in the past focused on searching for a few functional variants. Although such GWAS studies have met some successes in identifying a number of variants, these variants account for only a small fraction of the total trait variation and their functional roles typically remain unclear. The MGD predicts that complex diseases may be caused by excess genetic noise over a threshold and may serve to prevent infinite increase in genetic diversity. Complex traits evolved as a result of suppressing genetic noises and hence should be susceptible to damage by excess noises. Also, insufficient amount of genetic diversity may hurt adaptive capacities such as immunity. The quantitative variations in a complex trait may correlate with the amount of genetic variations.

Recent results have shown the expected that higher MAC or noises correlates with higher lung cancer incidence in mice and humans [6,9]. Also, Parkinson's disease patients have higher MAC than controls and a selected set of ~37000 minor alleles can predict 2% of Parkinson's patients [8]. An efficient method of identifying target genes of complex traits has been established using the MAC concept [11]. MAC may control traits by regulating a set of target genes whose expressions are associated with both MAC and traits [11]. The quantitative nature of complex traits appear to be related to MAC and also to the degree of mitochondrial and nuclear genome compatibility, which is unexpected if most genome sequence are neutral or junks [10].

There is also new evidence for the inverse relationship between genetic diversity and epigenetic complexity. Individual animals with higher MAC or genetic variation are less able to maintain a stable epigenetic memory [7].

*Speciation.* The genetic mechanisms of speciation remain better understood, which really is the ultimate challenge in evolutionary studies. A genic view popular at the moment suggests adaptive selection of small sets of genes as being responsible for reproductive isolation [87,88]. In contrast, classic biological species concept of Mayr suggests reproductive isolation as genome wide phenomenon[89], and such a non-genic view is more than a century old [90,91]. As explained by Forsdyke: "In his theory of 'physiological selection', Romanes postulated germline 'collective variations' that accumulate in certain members of a species; these members are thus 'physiological complements' producing fertile offspring when mutually crossed, but sterile offspring when crossed with others. Unlike Darwin's natural selection, which secured reproductive isolation of the fit by elimination of the unfit, physiological selection postulated variations in the reproductive system that were not targets of natural selection; these sympatrically isolated the fit from the fit, leaving two species where initially there had been one. Bateson approved of physiological selection… and postulated a 'residue', distinct from genes, that might affect gene flow between organisms and so originate species. The reproductive isolation of the parents of a sterile hybrid was due to two complementing non-genic factors (the 'residue') separately introduced into the hybrid by each parent" [91]. Forsdyke has further suggested base composition as the non-genic factor that could influence homologous chromosome pairing [91]. In support of this view, derived species do appear to have more A and T nucleotides [92].

The Romanes–Bateson viewpoint is remarkably consistent with the MGD thesis and even the terms used are highly similar. Unaware of this century old and much ignored viewpoint until a recent internet exchange with Forsdyke (http://sandwalk.blogspot.com/2015/04/does-natural-selection-constrain.html), we have used the term 'collective effects' in several of our recent papers [6-11]. A key novel point of the MGD thesis is to emphasize the role of internal system or physiology in determining the MGD of a species, which may be termed physiological selections.

Different individuals with different MAC in a population may segregate into different groups with individuals of similar MAC and hence similar traits clustering together into one group. Higher MAC is expected to show base composition differences from lower MAC, because minor alleles as well as new mutations are enriched in A and T nucleotides. Speciation during microevolution (as defined in Fig. 1b) may result from the original separation of individuals based on their MAC values. A prediction of this non-genic view is that organisms with large MGD such as the bacteria kind should have much greater potential to subdivide into multiple species than those with lower MGD, since low diversity may not give rise to enough genome wide base composition differences to interfere with homologous chromosome pairing. Especially if Nei's theory of survival of the niche-filling variants is correct, organisms with large MGD or more genetic variants should be able to fill many niches and hence evolve via micro-evolutionary mechanisms into more species of similar complexity [75]. In contrast, speciation according to the genic view should not correlate with MGD or complexity. Even a casual read of the diversity of species on Earth would find a general trend of an inverse relationship between complexity of the clade and number of species within the clade.

**Summary**

Compare the two approaches to study evolution. One ignores complexity. The other engages it even if imperfectly. Which approach is better or has a chance to eventually succeed? If one thinks human is no more complex than HIV or bacteria, he surely would pick approach one. But then he is clearly not at the level of Aristotle in terms of intuitive power and really should find it hard to succeed in the thinking profession or the hard science, the foundation of which is intuition or axiom. Indeed, the approach of ignoring complexity has been given all it ever needs to succeed for the past century and yet still has not delivered any insight to the old riddle of genetic diversity.

Lewontin has observed in 1974: "How can such a rich theoretical structure as population genetics fail so completely to cope with the body of fact? Are we simply missing some critical revolutionary insight that in a flash will make it all come right, as the Principle of Relativity did for the contradictory evidence on the propagation of light? Or is the problem more pervading, more deeply built into the structure of our science? I believe it is the latter." [5]. We however believe it is the former, and in any event, believing in the latter may never lead to a constructive solution. Hard sciences like physics are built by mathematics and axioms. While population genetics may not be short of mathematics, the foundations for its mathematics are all man-made assumptions that are all self-evidently untrue or only true in a very limited sense. Most counter-intuitive of all, the MET has long ignored complexity and treated evolution from one type of bacteria to another type the same as from bacteria to human. The introduction of a couple of axioms as in the MGD hypothesis appears sufficient to account for the major phenomenology and to absorb all of the proven virtues of MET. With now a solid foundation finally established, we anticipate a bright future for the field in producing precise and certain results that could rival any hard sciences. Such evolutionary insights may also stimulate progresses in other fields of biomedical sciences.

**Acknowledgements:**

We thank Donald Forsdyke for critical reading of the manuscript. This work was supported by the National Natural Science Foundation of China (Grant No. 81171880) and the National Basic Research Program of China (Grant No. 2011CB51001).

**Figure legends:**

**Figure 1. Schematic representation macroevolution (a) and microevolution (b).**
Yellow color represents allowed or tolerable mutant sites in a sequence. Orange color represents sites where actual mutations have occurred.

# A

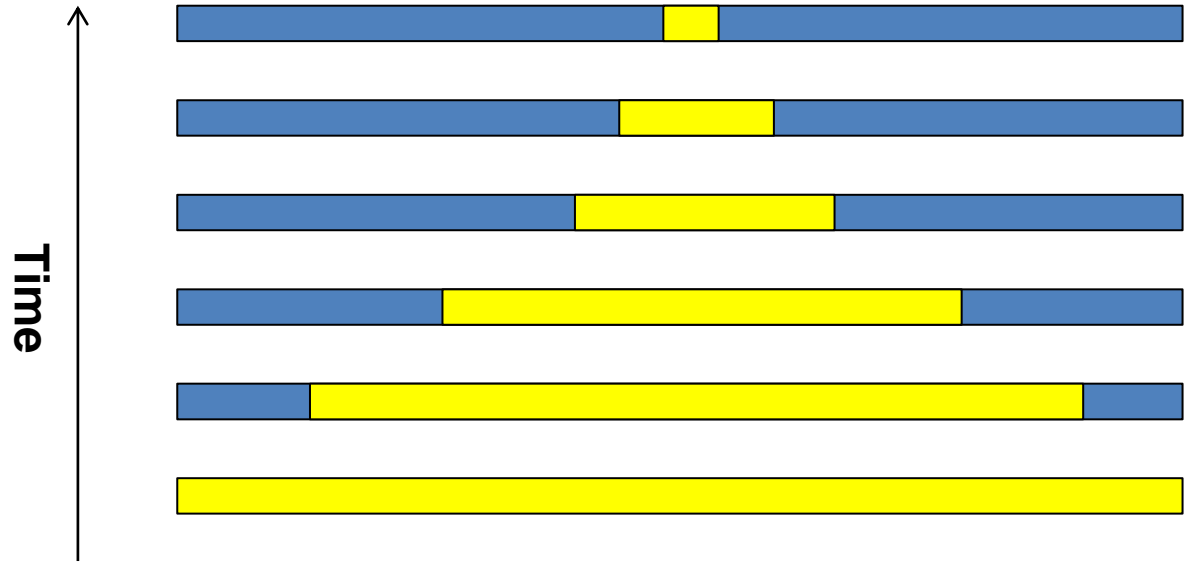

# B

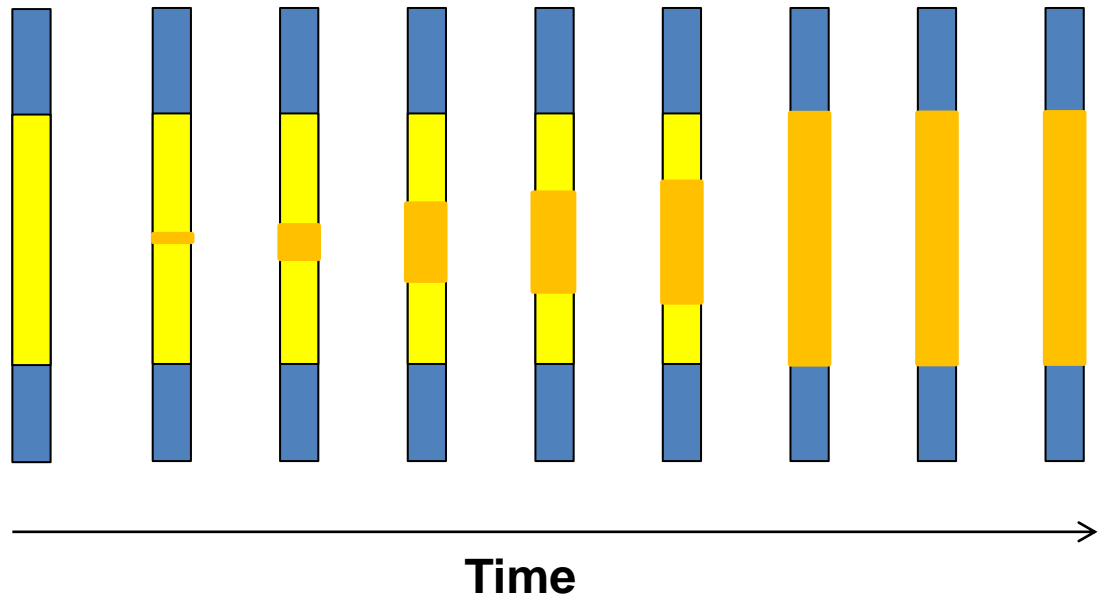

Time